\documentstyle[11pt,paspconf,epsf]{article}
\newcommand\simlt{\lower.5ex\hbox{$\; \buildrel < \over \sim \;$}}
\newcommand\simgt{\lower.5ex\hbox{$\; \buildrel > \over \sim \;$}}

\begin{document}

\title{Pressure Ionization Instability: Connection between Seyferts
and GBHCs}

\author{Sergei Nayakshin} 

\affil{Department of Physics, the University of Arizona, Tucson, AZ, 85721}

\begin{abstract}
  Spectrum of Seyfert 1 Galaxies (S1G hereafter) is very similar to
  that of several Galactic Black Hole Candidates (GBHCs) in their hard
  state, suggestive that both classes of objects have similar physical
  processes. However, recent work has shown that reprocessing features
  make the patchy corona-disk model (PCD model: the best current
  explanation of S1G spectrum) problematic for GBHCs. To address the
  similarities and differences in spectrum of Seyferts and GBHCs, we
  consider the structure of the ionized X-ray skin near an active
  magnetic flare. We show that the X-ray skin is subject to a thermal
  instability, similar in nature to the well known ionization
  instability of quasar emission line regions. Due to the much higher
  ionizing X-ray flux in GBHCs, the only stable solution for the upper
  layer of the accretion disk is that in which it is highly ionized
  and is at the Compton temperature ($\sim $ few keV). We show that
  this accounts for the difference in spectrum of GBHCs and S1G.  In
  addition, same instability, applied to S1G, leads to the X-ray skin
  temperature $T\sim 1-3\times 10^5$ Kelvin, which then may explain
  the observed spectral shape of BBB.
\end{abstract}

\keywords{accretion disks,magnetic fields,radiation mechanisms:
non-thermal,instabilities,radiative transfer,atomic processes}

\section{Introduction}\label{sect:intro4}

The X-ray spectra of S1G and GBHCs indicate that the reflection and
reprocessing of incident X-rays into lower frequency radiation is an
ubiquitous and important process (Pounds et al. 1990, Nandra \& Pounds
1994; Zdziarski et al. 1996). It is generally believed that the
universality of the X-ray spectral index in S1G ($\Gamma\simeq 1.9$)
may be attributed to the fact that the reprocessing of X-rays within
the disk-corona of the two-phase model leads to an electron cooling
rate that is roughly proportional to the heating rate inside the
active regions (AR) where the X-ray continuum originates (Haardt \&
Maraschi 1991, 1993; Haardt, Maraschi \& Ghisellini 1994; Svensson
1996). Although the X-ray spectra of GBHCs are similar to that of
Seyfert galaxies, they are considerably harder (most have an intrinsic
power-law index of $\Gamma \sim 1.5-1.7$), and the reprocessing
features are less prominent (Zdziarski et al. 1996). Dove et
al. (1997) recently showed that a Rossi X-ray observation of Cygnus
X-1 shows no significant evidence of reflection features. The
relatively hard power law and the weak reprocessing/reflection
features led Dove et al.  (1997, 1998), Gierlinski et al. (1997) and
Poutanen, Krolik \& Ryde (1997) to conclude that the PCD model does
not apply to Cygnus X-1. This conclusion is sensitive to the
assumption that the accretion disk is relatively cold, such that $\sim
90$\% of the reprocessed coronal radiation is re-emitted by the disk
as thermal radiation (with a temperature $\sim 150$ eV). To test
validity of this assumption, we extend earlier work of Nayakshin \&
Melia (1997), who investigated the X-ray reflection process in AGNs
assuming that the ARs are magnetic flares above the disk, on the case
of GBHCs.

\section{Thermal Instability of the Transition Region}

\begin{figure*}
\plotfiddle{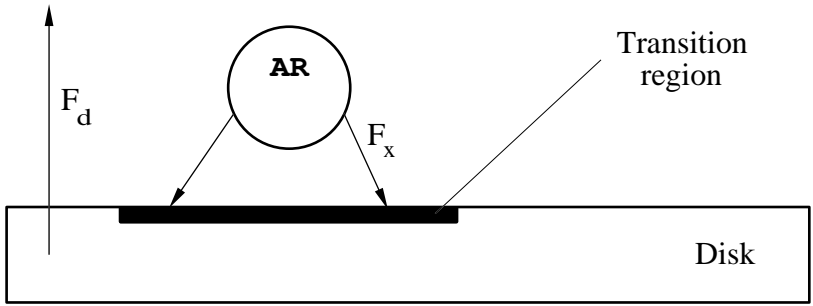}{100pt}{0}{50}{50}{-150}{-220}
\caption{The geometry of the active region (AR, the X-ray source) and
the transition layer. Magnetic fields, confining AR and supplying it
with energy are not shown. Transition region is defined as the upper
layer of the disk with Thomson depth of $\sim$ few, where the incident
X-ray flux $F_{\rm x}$ is substantially larger than the intrinsic disk
flux $F_{\rm d}$.}
\label{fig:active_region}
\end{figure*}

The relevant geometry is shown in Figure (\ref{fig:active_region}).
Since the flux of the ionizing radiation from the active region is
rapidly declining with distance away from the flare, only the gas near
the active regions (with a radial size $\sim$ a few times the size of
the active region) may be highly ionized. To distinguish these
important X-ray illuminated regions from the ``average'' X-ray skin of
the accretion disk (i.e., far enough from active magnetic flares), we
will refer to these as transition layers (or regions).  We will only
consider the structure of the cold disk in the transition layer and
only solve the radiation transfer problem for these regions as well,
because this is where most reprocessed coronal radiation will take
place.

The compactness parameter of the active region, $l$, is defined as
\begin{equation}
l\equiv {F_{\rm x}\sigma_T \Delta R\over m_e c^3},
\label{compact}
\end{equation}
and is expected to be larger than or of order of unity (e.g., Poutanen
\& Svensson 1996, Poutanen, Svensson \& Stern 1997, Chapter 2 in
Nayakshin 1998b). Here, $F_{\rm x}$ is the X-ray illuminating flux
from the flare, $\sigma_T$ is the Thomson cross section, and $\Delta
R$ is the size of the active region $\Delta R$, which is thought to be
of the order of the accretion disk height scale $H$ (e.g., Galeev et
al.  1979).  Inverting this definition, one gets an estimate for
$F_{\rm x}$ for a given compactness parameter. Due to space
limitations, we just mention that one can show that (1) during the
flare, the ionizing X-ray flux $F_{\rm x}$ substantially exceeds the
disk thermal flux $F_{\rm disk}$ in parameter space appropriate for
both S1Gs and GBHCs; (2) the X-radiation ram pressure, $F_{\rm x}/c$
is much larger than the gas pressure in the disk atmosphere (Nayakshin
1998b, Nayakshin \& Dove 1998, paper I \& II hereafter). Under these
conditions, the ionizing radiation ram pressure compresses the disk
atmosphere (in the transition layer only, of course) so that the gas
pressure there matches the ram pressure, i.e., $P \simlt F_{\rm x}/c$.

Thermal instability was discovered by Field (1965) for a general
physical system. He introduced the ``cooling function'' $\Lambda_{\rm
net}$, defined as difference between cooling and heating rates per
unit volume, divided by the gas density $n$ squared. Energy equilibria
correspond to $\Lambda_{\rm net} = 0$. He argued that a physical
system is usually in pressure equilibrium with its surroundings. Thus,
any perturbation of the temperature $T$ and the density $n$ of the
system should occur at a constant pressure. The system is unstable
when
\begin{equation}
\left({\partial \Lambda_{\rm net}\over \partial T}\right)_{P} < 0,
\label{field}
\end{equation}
since then an increase in the temperature leads to heating increasing
faster than cooling, and thus the temperature continues to increase.
Similarly, perturbation to a lower $T$ will cause the cooling to
exceed heating, and $T$ will continue to decrease.

In ionization balance studies, it turns out convenient to define two
parameters. The first one is the ``density ionization parameter''
$\xi$, equal to (Krolik, McKee \& Tarter 1981) $\xi = 4\pi F_{\rm
x}/n$.  The second one is the ``pressure ionization parameter'',
defined as $\Xi = F_{\rm x}/(2 c n k T) \equiv P_{\rm rad}/P$, where
$P$ is the gas pressure. This definition of $\Xi$ is the one used in
the ionization code XSTAR (see below), and is different by factor
$2.3$ from the original definition of Krolik et al. (1981), who used
the hydrogen density instead of the electron density. In papers I \&
II, we showed that Fields instability criterion is equivalent to the
following condition (see also Krolik et al. 1981):
\begin{equation}
\left( {d\Xi\over d T}\right)_{\Lambda_{\rm net}=0}\, < 0
\label{fcond}
\end{equation}

We now apply the X-ray ionization code XSTAR, written by T. Kallman
and J.  Krolik, to the problem of the the transition layer. A truly
self-consistent treatment would involve solving radiation transfer in
the optically thick transition layer, and, in addition, finding the
distribution of the gas density in the transition layer that would
satisfy pressure balance. Since radiation force acting on the gas
depends on the opacity of the gas, this is a difficult non-linear
problem. We defer such a detailed study to future work, and simply
solve (using XSTAR) the local energy and ionization balance for {\em
an optically thin layer} of gas in the transition region. We assume
that the ionizing spectrum consists of the incident X-ray power law
with the energy spectral index typical of GBHCs in the hard state,
i.e., $\Gamma = 1.5-1.75$, exponentially cutoff at 100 keV, and the
blackbody spectrum at temperature $T_{\rm min}$ with flux equal to the X-ray
flux. We include the former component to mimic the spectrum reflected
from the cold disk below the transition layer (usually as much as
$80-90$ \% of the reflected flux comes out as the cold black body
emission, see \S 3).

When applying the code, one should be aware that it is not possible
for the transition region to have temperature lower than the effective
temperature of the X-radiation, i.e., $T_{\rm min} = (F_{\rm
x}/\sigma)^{1/4}$.  The reason why simulations may give temperatures
lower than $T_{\rm min}$ is that in this parameter range XSTAR
neglects certain de-excitation processes, which leads to an
overestimate of the cooling rate for $T\sim T_{\rm min}$ (\'Zycki et
al. 1994; see their section 2.3).  In the spirit of one zone
approximation for the transition layer, we use an average X-ray flux
$\langle{F_{\rm x}}\rangle$ as seen by the transition region, which we
parameterize as $\langle{F_{\rm x}}\rangle = 0.1 F_{\rm x}/q_1$, where
$q_1 = q/10$, and $q$ is a dimensionless number of order 10 (see
figure \ref{fig:active_region}; $F_{\rm x}$ is the X-ray flux {\em at}
the active region). Nayakshin (1998b) shows that
\begin{equation}
T_{\rm min}\simeq 5.0 \times 10^6\; l^{1/4} \, q_1^{-1/4}\, \left({\dot{m}\over
0.05}\right)^{-1/20}\, \alpha^{1/40} \, M_1^{-9/40} \,
\left[1-f\right]^{-1/40}
\label{tmincyg}
\end{equation}
where $l\gg 0.01$ is the compactness parameter, $\dot{m}$ is the
dimensionless accretion rate, $M_1\equiv M/10\,M_{\odot}$, $f$ is the
fraction of power supplied from the disk to corona (see Nayakshin
1998a) and $\alpha$ is the viscosity parameter.

\begin{figure*}
\plotfiddle{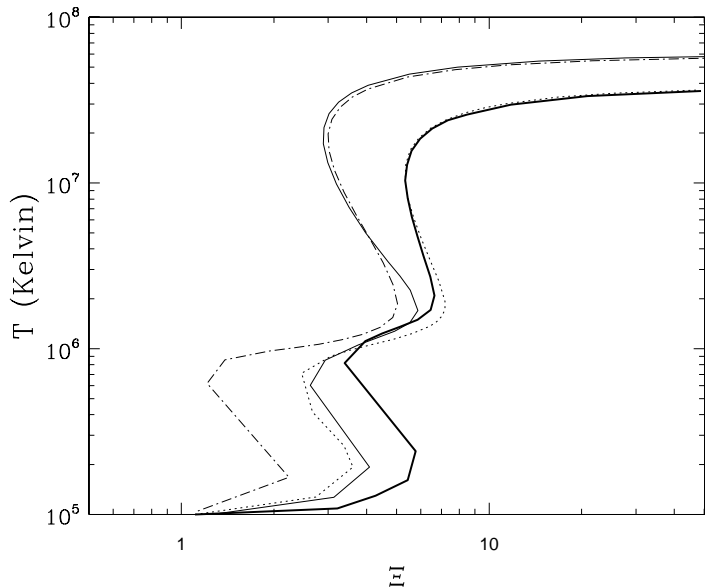}{150pt}{0}{50}{50}{-150}{-150}
\caption{Gas temperature versus pressure ionization parameter $\Xi$ --
  for GBHCs. Values of the parameters are: $\Gamma =$ 1.5, 1.75, 1.75,
  1.7 and $k T_{\rm min} = $ 200, 100, 200, 400 eV, corresponding to
  the fine solid, thick solid, dotted and dash-dotted curves,
  respectively. The ionization equilibrium is unstable when the curve
  has a negative slope. Note that there exist no solution for $T$
  below $T_{\rm min}$.}
\label{fig:scurve}
\end{figure*}

Figure \ref{fig:scurve} shows results of our calculations for several
different X-ray ionizing spectra. A stable solution for the transition
layer structure will have a positive slope of the curve, and also
satisfy the pressure equilibrium condition. As discussed earlier, $P
\leq F_{\rm x}/c$ (i.e., $\Xi\geq 1$). In addition, if the gas is
completely ionized, the absorption opacity is negligible compared to
the Thomson opacity. Because all the incident X-ray flux is eventually
reflected, the net flux is zero, and so the net radiation force is
zero.  In that case $P$ adjusts to the value appropriate for the
accretion disk atmosphere in the absence of the ionizing flux (see
also Sincell \& Krolik 1996), i.e., $\Xi\gg 1$. Thus, the upper branch
of the ionization equilibrium curve, where the transition layer is at
the Compton equilibrium temperature, is stable, because the ionization
curve has a positive slope and the large values of $\Xi$ are
physically allowed.

In addition to the Compton equilibrium state, there is a smaller
stable region for temperatures in the range between $100$ and $200$
eV. The presence of this region is explained by a decrease in {\em
heating}, rather than an increase in cooling (cf. equation \ref{field}
and recall $\Lambda_{\rm net}=$ cooling -- heating). The X-ray heating
decreases in the temperature range $100-200$ eV with increasing $T$
because of consequent destruction (ionization) of ions with
ionization energy close to $k T$. This is highly unlikely that the
transition region will stabilize at the temperature $100$ -- $200$ eV,
because the effective temperature $T_{\rm min}$ is at or above this
temperature range. Further, Nayakshin (1998b) considered the effects
of the radiation pressure in the transition layer more accurately by
computing the gas cross sections to the incident and reprocessed
fluxes. He shows that the stable state with $kT\sim 100-200$ eV is
forbidden on the grounds of the pressure equilibrium.

Rounding this discussion up, we believe that the only stable
configuration available for the transition layer of GBHCs in the hard
state is the one at the local Compton temperature. Future work should
concentrate on finding not only the exact value of $\tau_x$, but the
exact distribution of gas temperature, density and ionization state in
the atmosphere of the accretion disk as well. For now, however, we
will treat $\tau_x$ as a free parameter and numerically investigate
the ramifications of the transition layer on the spectrum of escaping
radiation and the physical properties of the corona.

\section{``Three-Phase'' Model for GBHCs}\label{sect:threephase}

To explore how the structure of the ionized transition region affects
the X-ray spectrum from magnetic flares, we computed the X-ray
spectrum from a magnetic flare above the transition layer with a range
of $\tau_{\rm trans}$.  The gas in the active region is heated
uniformly throughout the region and is cooled by the Compton
interactions with radiation re-entering the active region from below.
Even though the geometry of the AR is probably closer to a sphere or a
hemisphere than a slab, we shall adopt the latter for numerical
convenience, neglecting the boundary effects.  Experience has shown
that spectra produced by Comptonization in different geometries are
usually qualitatively similar (i.e., a power-law plus an exponential
roll-over), and it is actually the fraction of soft photons entering
the corona that accounts for most of the differences in the various
models, because it is this fraction that affects the AR energy
balance. To crudely take geometry into account, we permit only a part
of the reprocessed radiation to re-enter the corona, and fix this
fraction at $0.5$ (cf. Poutanen \& Svensson 1996). The Thomson optical
depth of the corona is fixed at $\tau_{\rm c} = 0.7$. We employ the
Eddington (two-stream) approximation for the radiative transfer in
both the AR and the transition layer.

The disk below the flare is broken into two regions: (i) the
completely ionized transition region, situated on the top of (ii) the
cold accretion disk, which emits blackbody radiation at a specified
temperature. We model the transition layer as being one
dimensional. The X-radiation enters the transition region through its
top. In this region, the only process taken into account is the
Compton scattering. After being down-scattered, the X-radiation is
``incident'' on the cold accretion disk from the bottom of the
transition layer.  The incident spectrum is reflected in the standard
manner (Magdziarz \& Zdziarski 1995). The total radiation spectrum
re-entering the transition layer from below is the sum of the
reflection component and the blackbody component due to the disk
thermal emission, which is normalized such that the incident flux from
the transition region is equal to the sum of the fluxes from the
reflection component and the blackbody.  The optically thick cold disk
is held at a temperature $T_{\rm bb} = 2.4 \times 10^6$ Kelvin.

The observed spectrum consists of the direct component, emerging
through the top of the AR, and a fraction of the reflected radiation
that emerges from the transition layer and does not pass through the
corona on its way to us (see Fig. 1). This fraction is chosen to be
0.5 as well.  Physically, it accounts for the fact that, as viewed by
an observer, a part of the transition region itself is blocked by the
active region. The overall setup of the active region - disk
connection is very similar to the one used by Poutanen \& Svensson
(1996), except for the addition of transition layer on the top of the
cold disk.

\begin{figure*}
\plotfiddle{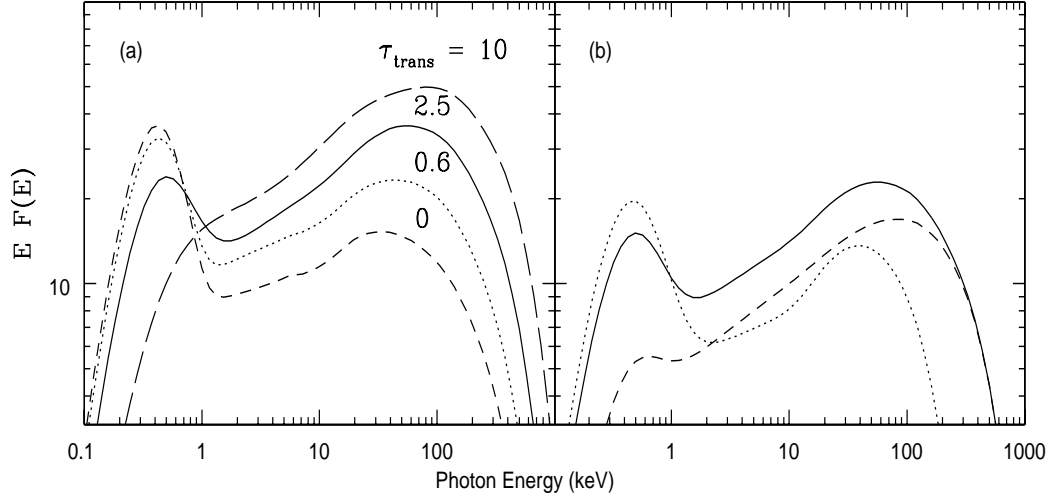}{150pt}{-90}{55}{65}{-230}{230}
\caption{(a) Resulting spectrum from the patchy corona disk model as a
function of the Thomson optical depth $\tau_{\rm trans}$ of the
transition layer. Notice that higher values of $\tau_{\rm trans}$ lead
to harder spectrum with the disk blackbody component getting
progressively smaller. (b) Decomposition of the total spectrum on its
constituents.  See text for details.}
\label{fig:sequence_of_spectra}
\end{figure*}

Figure (\ref{fig:sequence_of_spectra}a) shows the ``observed''
spectrum for several values of $\tau_{\rm trans}$: $0$, $0.6$, $2.5$,
and $10$.  It can be seen that the spectrum hardens as $\tau_{\rm
trans}$ increases and the fraction of energy in reprocessed (soft)
component below $\sim 2$ keV decreases, which can be understood by
noting that a larger fraction of the photons from the AR is reflected
before they have a chance to penetrate into the cold disk where the
blackbody component is created.  In Figure (3b) we show the components
that contribute to the overall spectrum for $\tau_{\rm trans} =
3$. The solid, dashed and dotted curves show the total spectrum, the
AR intrinsic spectrum and the reprocessed spectrum (emerging from the
top of the transition layer). Notice that the reprocessed component
has about equal amount of power below and above 2 keV, whereas the
usual division of power in the reflected spectrum ({\it from a neutral
reflector}) is $80-90$ in the soft and $20-10$ \% in the hard
components, correspondingly (e.g., Magdziarz \& Zdziarski 1995). This
is the most profound difference between our calculations and those of
previous workers, who assumed that the disk boundary is infinitely
sharp, so that there is no transition layer between the AR and the
cold disk.

Gierlinski et al. (1997) have attempted to fit broad-band spectrum of
Cyg X-1 with active regions above a cold accretion disk, and showed
that the most difficult issue for the two-phase model is the too small
observed amount of the reprocessed soft X-radiation. For example,
Zheng et al.  (1997) shows that Cyg X-1 luminosity in the hard state
below 1.3 keV is about $5\times 10^{36}$ erg/s, whereas the luminosity
above 1.3 keV is $\sim 3-4\times 10^{37}$ erg/s. This is impossible in
the context of the simple PCD model, since about half of the
X-radiation impinge on the cold disk and get reprocessed into the
blackbody radiation. Accordingly, the minimum luminosity in soft
X-rays below 1.3 keV should be about that of the hard component.
However, we find that, with the advent of the transition layer, the
combined power below $2$ keV accounts for only $25\;\%$ of the total
for $\tau_{\rm trans} = 3$. Further, notice that the spectra are
correspondingly harder in X-rays, which explains why GBHCs spectra are
harder than those of typical S1G. In paper I we address the other
reprocessing features (e.g., the iron line and anisotropy break) and
show that the theory predictions are consistent with Cyg X-1 spectrum.
In paper II, we apply the non-linear Monte-Carlo routine (for details
of the routine and geometry, see Dove, Wilms, \& Begelman 1997), and
demonstrate that our results obtained with the simpler Eddington
approximation code hold true.

\section{The Pressure Ionization Instability for AGN}
\label{sect:inst_agn}

We now discuss the thermal instability of the surface layer for AGN.
The most important distinction from the GBHC case is the much higher
mass of the AGN, and thus the ionizing X-ray flux is smaller by $\sim
7$ orders of magnitude (since $F_{\rm x}\propto L/R^2 \propto \dot{m}
L_{\rm Edd} /R_s^2\propto \dot{m} M^{-1}$). The minimum X-ray skin
temperature is again approximated by setting the blackbody flux equal
to the incident flux. The gas pressure dominated solution gives (paper
I)
\begin{equation}
T_{\rm min}\simeq 1.5\times 10^5\, l^{1/4} \,\alpha^{1/40} \,M_8^{-9/40}
\left[{\dot{m}\over 0.005}\right]^{-1/20} \,\left(1-f\right)^{-1/40} \,
\left({q\over 10}\right)^{-1/4}\,
{\rm ,}
\label{tminsg}
\end{equation}
whereas the radiation-dominated one yields
\begin{equation}
T_{\rm min}\simeq 1.24\times 10^5\, l^{1/4} \,M_8^{-1/4}
\left[{\dot{m}\over 0.005}\right]^{-1/4} \,\left(1-f\right)^{-1/4}\,
\left({q\over 10}\right)^{-1/4}\,
\label{tminsr}
\end{equation}

These estimates show our main point right away: the lower X-ray flux
density in AGN may allow the transition layer to saturate at either
the cold equilibrium state or the ``island'' state with $T\sim 100-
200$ eV, whereas that was not possible for GBHCs. To investigate this
idea, we ran XSTAR as described in \S 2, but for parameters
appropriate for an AGN transition layer. The X-rays illuminating the
transition region are assumed to mimic the typical Seyfert hard
spectra, i.e., a power-law with photon index $\Gamma = 1.9$ and the
exponential roll-over at 100$-$200 keV range. We also add the
reflected blackbody component as described in \S 3.

\begin{figure*}[t]
\plotfiddle{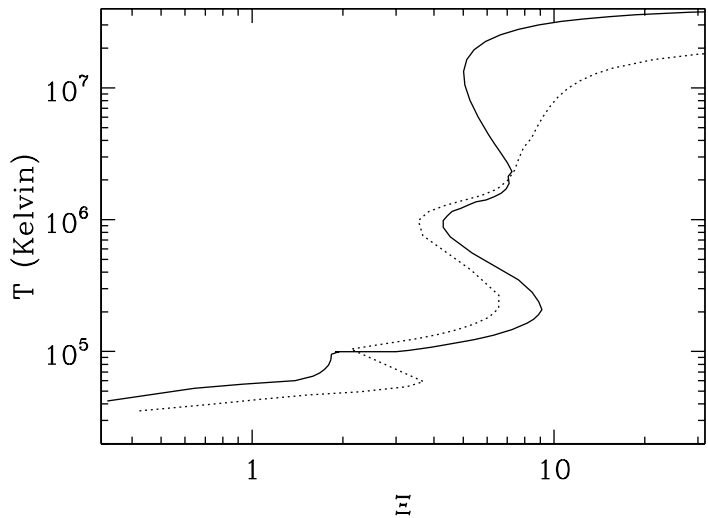}{150pt}{0}{50}{50}{-160}{-160}
\caption{Same as Figure (2), but for an AGN transition layer.}
\label{fig:agn1}
\end{figure*}

We show results of two such simulations in Figure
(\ref{fig:agn1}). The solid curve corresponds to $T_{\rm min} = 6$ eV and
the rollover energy of $100$ keV, while for the dotted curve $T_{\rm min} =
12$ eV and the rollover energy of $200$ keV.  As explained earlier,
XSTAR produces inaccurate results below $T\sim T_{\rm min}$, so that these
regions of the ionization equilibrium curve should be disregarded.
Notice that the ``cold'' equilibrium branch, i.e., the region with
$T\sim 10^5$ K is broader in terms of $\Xi$ than the island
state. Further, preliminary more detailed pressure equilibrium
considerations (paper I) show that the island state is unlikely to
satisfy the pressure equilibrium, so that the two truly stable
solutions for the transition layer in AGN are the cold stable state
with $kT \sim 10-30$ eV and the hot Compton equilibrium state, which
we already discussed for GBHCs in \S 3.

In addition, the Rosseland mean optical depth to the UV emission is of
order 1 to few for the temperature range $kT \sim 10-30$ eV (paper I).
As one can check using Field (1965) stability criterion, the
transition layer radiating via blackbody or modified emission is
thermally stable. This consideration adds weight to our optically thin
calculations in that the cold state of the transition layer in AGN
disks should be unquestionably stable.

It is interesting to note that the reflection component and the
fluorescent iron line that are always present in the spectra of
radio-quiet Seyfert Galaxies (e.g., Gondek et al. 1996, Zdziarski et
al. 1996, George \& Fabian 1991) can be best fitted with a neutral or
weakly ionized reflector. From work of Matt, Fabian \& Ross (1993,
1996) and Zycki et al. (1994), it is known that the ionization
parameter $\xi\simlt 100$ is required to fit S1G data.  The cold
stable solution found here corresponds to $\xi$ ranging from few tens
to $\sim 200$ (paper I), thus being consistent with observations of
X-ray reflection and iron lines in Seyferts.

\section{The Origin of the Big Blue Bump (BBB) in Seyferts}
\label{sect:obbb}

In recent years, there has been considerable progress in observations
of the BBB (e.g., Walter \& Fink 1993; Walter et al. 1994; Zhou et
al. 1997). It was found that the observed spectral shape of the bump
component in Seyfert 1's hardly varies, even though the luminosity $L$
(of the bump) ranges over 6 orders of magnitude from source to source.
This fact is uneasy to understand from the point of view of any {\it
disk} emission mechanism (see, e.g., Nayakshin 1998b, Chapter 5 \&
references there).

We believe that our theory of the ionization pressure instability may
offer a plausible explanation for the BBB emission. As our ionization
equilibria calculations show, there is no stable solution for the
transition region in the temperature range $\sim 3\times 10^5 \simlt T
\simlt 10^7$ Kelvin. Furthermore, temperatures below the effective
temperature of the X-ray radiation are also forbidden. If PCD model is
correct at all, $l\gg 0.01$. Preliminary estimates (Nayakshin 1998b)
show that $l\sim 0.1$ is required to explain soft X-ray part of
Cyg~X-1 spectrum. Since we believe it is the same physics of the PCD
model that explains both GBHCs and Seyferts, $l\sim 0.1$ yields $T_{\rm min}
\sim 10^5$ K (see equations \ref{tminsg} \& \ref{tminsr}) for
AGN. Thus, the only low temperature solution permitted by the
stability analysis for AGN with $M\sim 10^8\,M_{\odot}$ is the one with
temperature $1-3\times 10^5$ Kelvin. From our calculations, we also
found that the Rosseland mean optical depth to the UV emission is of
order 1 to few in the cold stable state. The radiation spectrum
produced by the transition layer will therefore be either a blackbody
spectrum, or a modified blackbody (with recombination lines as well,
of course). Since a moderately optically thick emission spectrum
saturates at photon energy of $\sim 2- 4 \times k T$, $T\sim 2 \times
10^5$ provides an excellent match to the observed roll-over energies
of $\sim 40-80$ eV (e.g., Walter et al. 1994).

The most attractive feature of this suggestion is that the temperature
of the BBB is fixed by atomic physics, in particular by the fact that
many atomic species have ionization potential close to 1 Rydberg
$\simeq 1.5 \times 10^5$ K, which may explain the fact that the BBB
shape changes so little from source to source. The stable temperature
range is independent of the number of magnetic flares, and so it is
independent of the X-ray luminosity of the source, as found by Walter
\& Fink (1993) and Walter et al.  (1994). Further, supplemented by our
theory of the division of power between the corona and the disk, the
pressure ionization instability can explain disappearance of the bump
for AGN more luminous than typical S1G, e.g., quasars from Zheng et
al. (1997) and Laor et al. (1997) samples (see Nayakshin 1998a and
paper I).

\section{Discussion}\label{sect:discussion}

By considering the irradiated X-ray skin close to an active magnetic
flare above a cold accretion disk, we have shown that the skin
equilibrium is in general unstable. Two stable states (one cold and
one hot) exist. For the case of GBHCs, we showed that the low
temperature equilibrium state is forbidden due to a high value of the
ionizing flux. Thus, the X-ray irradiated skin of GBHCs must be in the
hot equilibrium configuration, where the gas is at the local Compton
temperature ($k T\sim$ few keV). In an attempt to determine the
effects of this skin on the spectrum from a magnetic flare, we modeled
the ionization structure of the disk by assuming a completely ionized
layer with Thomson optical depth of $\sim$ few to be situated on the
top of the cold disk (cf. Fig. 1).

We found that the transition layer alters the reflected spectrum
significantly, and that it leads to GBHCs spectra being harder than
Seyfert 1 spectra for {\it same} parameters of magnetic flares.  We
also found that the highly ionized transition layer can account for
the disappearance/weakening of the reprocessing features, such as iron
line. We thus conclude that spectrum of GBHCs in their hard state is
consistent with PCD model when one takes into account the pressure
ionization instability discussed here.

Applying our results to AGN case, we found that, due to a
substantially lower ionizing flux as compared to GBHCs case, there
exists a stable solution for the transition layer in the temperature
range $T\sim 1-3 \times 10^5$ K. Thus, the reprocessed features are
expected to be characteristic of cold, almost neutral reflector, being
consistent with observations of AGN (e.g., Zdziarski et al. 1996). The
narrow range in the temperature of the transition region in S1G may
explain the observed roll-over energies in the BBB spectrum.

Commenting on the distinction of our work on the ionization structure
of the disk from extensive previous studies of this issue (e.g., Zycki
et al. 1994, Ross, Fabian \& Brandt 1996 and references there), we
note that the difference is caused by two factors: (1) previous
workers assumed that the corona is uniform and covers the whole disk,
whereas here we test the case of strongly localized emission from
magnetic flares, and (2), more importantly, previous studies fixed the
X-ray skin gas density at the disk mid-plane value, which, we note,
have little to do with the disk atmosphere density. The usual
statement that the density of radiation-dominated disks is
approximately constant is only correct as long as one stays deep
inside the disk, far from the surface (see, e.g., \S 2a and Fig. 11 of
Shakura \& Sunyaev 1973). Further, the pressure ionization instability
is not apparent in studies where the gas density is fixed to a
constant value, {\it regardless} of its value. As shown by Field
(1965), the thermal instability for the case with $n=$ const is always
weaker than it is for the case of a system in pressure equilibrium
(and it actually disappears in the given situation), which is
apparently the reason why this instability was not recognized before.

Thus, as far as we can see, observations of the hard state of the
GBHCs do not rule out magnetic flares as the source of X-rays, and
instead support this theory. We preliminary estimate that the observed
X-ray spectrum of Cyg~X-1 can be explained by the transition optical
depth of $\sim 3$, which is physically plausible, and that, apart from
the self-consistent difference in the structure of the transition
layer, same parameters for magnetic flares might be used in both AGN
and GBHCs to explain their spectra.

\acknowledgments The author is very thankful to the workshop
organizers for the travel support, and to F. Melia for support and
useful discussions in an early stage of this work.

\end{document}